\documentclass[conference]{IEEEtran}
\usepackage{amsfonts}
\usepackage{mathrsfs}
\usepackage{amssymb,amsmath}

\usepackage{algorithm}
\usepackage{algorithmic}
\usepackage{amsmath,amssymb,epsfig,graphics,subfigure}
\usepackage{theorem}
\usepackage{array,color}
\usepackage[compress]{cite}

\theoremheaderfont{\normalfont\bfseries}

\begin{document}

\title{Robust Transceiver Design for AF MIMO Relay Systems with Column Correlations}

\author{Chengwen Xing$^\dag$, Zesong Fei$^\dag$, Yik-Chung Wu$^\ddag$, Shaodan Ma$^\ddag$ and Jingming Kuang$^\dag$
   \\ $^\dag$School of Information and Electronics, Beijing Institute of Technology, Beijing, China
\\Email: \{xingchengwen\}@gmail.com \ \{zesongfei, jmkuang\}@bit.edu.cn \\
$^\ddag$Department of Electrical and Electronic Engineering, The University of Hong Kong, Hong Kong\\
Email: \{sdma,ycwu\}@eee.hku.hk
}

\maketitle

\begin{abstract}

In this paper, we investigate the robust transceiver design for dual-hop
amplify-and-forward (AF) MIMO relay systems with Gaussian distributed channel estimation errors. Aiming at maximizing the mutual information under imperfect channel state information (CSI), source precoder at source and forwarding matrix at the relay are jointly optimized. Using some elegant attributes of matrix-monotone functions, the structures of the optimal solutions are derived first. Then based on the derived structure an iterative waterfilling solution is proposed. Several existing algorithms are shown to be special cases of the proposed solution. Finally, the effectiveness of the proposed robust design is demonstrated by simulation results.

\end{abstract}

\section{Introduction}
\label{sect:intro}

\IEEEPARstart{C}ooperative communication is one of the key parts of the future communication protocols, as the deployment of relays can improve link equality, extend coverage range and mitigate inference. In general, there are various relay strategies which are casted into three main categories, i.e., amplify-and-forward (AF), decode-and-forward (DF) and compressed-and-forward (CF). Among these relaying strategies, AF strategy which has the lowest complexity  is most suitable for practical implementation.

It is also well-established that adopting multiple antennas has a potential to improve overall wireless system performance. In order to reap both benefits promised by cooperative communication and multi-input multi-output (MIMO) systems, linear transceiver design for AF MIMO relaying systems has been widely researched in \cite{Guan08,Mo09,Rong09,Medina07,Tang07}. Generally, speaking there are two main kinds of criteria for transceiver design: capacity maximization and mean-square-error (MSE) minimization. Joint design of relay forwarding matrix and destination equalizer for minimizing MSE is discussed in \cite{Medina07} and \cite{Guan08}. Furthermore, joint design of source precoder relay forwarding matrix and destination equalizer for minimizing MSE is investigated in \cite{Rong09}. The capacity maximization transceiver design has been discussed in \cite{Tang07,Rong09}.

\let\oldthefootnote\thefootnote
\renewcommand\thefootnote{}
\footnote{This research work was supported in part by Ericsson and Sino-Swedish IMT-Advanced and Beyond Cooperative Program under Grant No.2008DFA11780.}
\let\thefootnote\oldthefootnote
In most of previous works, channel state information (CSI) is assumed to be perfectly known. However, this assumption cannot be met in practice. Channel estimation errors are always inevitable and drastically degrades system performance. It is well known that robust designs can reduce or mitigate the negative effects introduced by imperfect CSI. This is also the motivation of our work.
In this paper, we jointly optimize source precoder matrix and relay forwarding matrix for mutual information maximization under channel estimation errors. Based on the properties of matrix-monotone functions, the optimal structure of robust transceivers is derived. Then, an iterative water-filling solution is proposed. Finally, the numerical result shows the performance advantage of the proposed robust design.

The following notations are used throughout this paper. Boldface
lowercase letters denote vectors, while boldface uppercase letters
denote matrices. The notation ${\bf{Z}}^{\rm{H}}$ denotes the
Hermitian of the matrix ${\bf{Z}}$, and ${\rm{Tr}}({\bf{Z}})$ is the
trace of the matrix ${\bf{Z}}$. The notation ${\bf{Z}}^{1/2}$ is the
Hermitian square root of the positive semi-definite matrix
${\bf{Z}}$, such that ${\bf{Z}}^{1/2}{\bf{Z}}^{1/2}={\bf{Z}}$ and
${\bf{Z}}^{1/2}$ is also a Hermitian matrix. For a rectangular diagonal matrix ${\boldsymbol \Lambda}$, ${\boldsymbol \Lambda} \searrow$ denotes the main diagonal elements are in decreasing order and ${\boldsymbol \Lambda} \nearrow$ denotes the main diagonal elements are in increasing order. For two Hermitian matrices, ${\bf{C}} \succeq
{\bf{D}}$ means that ${\bf{C}}-{\bf{D}}$ is a positive semi-definite
matrix. The symbol $\lambda_i({\bf{Z}})$ represents the $i^{\rm{th}}$ largest eigenvalue of ${\bf{Z}}$.

\section{System Model and Problem Formulation}
\subsection{Transmitted and Received Signals}
In our work, a dual-hop AF MIMO relay system is investigated, in which there
is one source with $N_S$ antennas, one relay with $M_{R}$ receive
antennas and $N_R$ transmit antennas, and one destination with $M_D$
antennas. Because of deep fading, the direct link between the source and destination is not taken into account. At the first hop, the source
transmits data to the relay. The received signal at the
relay is denoted as
\begin{align}
{\bf{x}}= {\bf{H}}_{sr}{\bf{P}}{\bf{s}}+{\bf{n}}_1
\end{align}where ${\bf{H}}_{sr}$ is the MIMO channel matrix between the source and
the relay, and ${\bf{P}}$ is the precoder matrix at the source. The
vector ${\bf{s}}$ is the $N \times 1$ data vector transmitted by the
source with the covariance matrix
${\bf{R}}_{s}=\mathbb{E}\{{\bf{s}}{\bf{s}}^{\rm{H}}\}={\bf{I}}_N$.
Furthermore, ${\bf{n}}_1$ is the additive Gaussian noise vector with
correlation matrix ${\bf{R}}_{n_1}=\sigma_{n_1}^2{\bf{I}}_{M_R}$.

At the relay, the received signal ${\bf{x}}$ is multiplied by a
forwarding matrix ${\bf{F}}$. Then the resultant signal is
transmitted to the destination. The received signal ${\bf{y}}$ at
the destination can be written as
\begin{equation}
\label{equ:signal} {\bf{y}} = {{\bf{H}}_{rd} {\bf{F}}
{\bf{H}}_{sr}{\bf{P}}{\bf{s}}}  + {{\bf{H}}_{rd} {\bf{F}}{\bf{n}}_1
} + {\bf{n}}_2,
\end{equation}where ${\bf{H}}_{rd}$ is the MIMO channel matrix between the relay
and the destination, and ${\bf{n}}_2$ is the additive Gaussian noise
vector at the second hop with covariance matrix ${\bf{R}}_{n_2}=\sigma_{n_2}^2{\bf{I}}_{M_D}$.
In order to guarantee the transmitted data ${\bf{s}}$ can be
recovered at the destination, it is assumed that $N_S$, $M_R$,
$N_R$, and $M_D$ are greater than or equal to $N$ \cite{Guan08}.

When channel estimation errors are taken into account, the dual-hop channels read as
\begin{align}
{\bf{H}}_{sr}={\bf{\bar H}}_{sr}+{\Delta}{\bf{H}}_{sr}, \ \ {\bf{H}}_{rd}={\bf{\bar H}}_{rd}+{\Delta}{\bf{H}}_{rd},
\end{align}where ${\bar{H}}_{sr}$ and ${\bar{H}}_{rd}$ are the channel estimates and ${\Delta}{\bf{H}}_{sr}$ and ${\Delta}{\bf{H}}_{rd}$ are the corresponding estimation errors with zero-mean Gaussian distributed entries. Additionally, the estimation errors are independent with each other as the channels are separately estimated. Referring to estimation errors, the following widely used Kronecker structure is adopted  \cite{Zhang08} \cite{Ding10}
\begin{align}
{\Delta}{\bf{H}}_{sr}={\boldsymbol \Sigma}_{sr}^{1/2}{\bf{H}}_{W,sr}{\boldsymbol \Psi}_{sr}^{1/2} \ \ {\Delta}{\bf{H}}_{rd}={\boldsymbol \Sigma}_{rd}^{1/2}{\bf{H}}_{W,rd}{\boldsymbol \Psi}_{rd}^{1/2},
\end{align} where the entries of ${\bf{H}}_{W,sr}$ and ${\bf{H}}_{W,rd}$ are identical and independent distributed (i.i.d.) with zero mean and unit variance. The column correlation matrices (${\boldsymbol \Psi}_{sr}$ and ${\boldsymbol \Psi}_{rd}$) and the row correlation matrices   (${\boldsymbol \Sigma}_{sr}$ and ${\boldsymbol \Sigma}_{rd}$) are determined by training sequences and channel estimators \cite{Xing1012}. To the best of the authors' knowledge, for a general case even for a point-to-point MIMO system, there is no closed-form solution. In this paper, we focus on the case with column correlations only i.e.,
\begin{align}
{\boldsymbol \Sigma}_{sr}=\alpha_1{\bf{I}}, \ \ {\boldsymbol \Sigma}_{rd}=\alpha_2{\bf{I}},
\end{align}as this case corresponds to a practical linear minimum mean square error (LMMSE) channel estimator \cite{Xing1012}.

\subsection{Problem Formulation}

At the destination, a linear equalizer ${\bf{G}}$ is adopted to
detect the data vector ${\bf{s}}$. The mean-square-error (MSE)
matrix is $\mathbb{E}\{({\bf{G}}{\bf{y}}-{\bf{s}})({\bf{G}}{\bf{y}}-{\bf{s}})^{\rm{H}}\} $ , where the expectation is taken with respect to random data, channel estimation errors, and noise.  In \cite{Xing10}, it is shown that
\begin{align}
\label{MSE_final} &\mathbb{E}\{({\bf{G}}{\bf{y}}-{\bf{s}})({\bf{G}}{\bf{y}}-{\bf{s}})^{\rm{H}}\}\nonumber \\& ={\bf{G}}({\bf{\bar
H}}_{rd}{\bf{F}}{\bf{R}}_{\bf{x}}{\bf{F}}^{\rm{H}} {\bf{\bar
H}}_{rd}^{\rm{H}}+{\bf{K}}_2){\bf{G}}^{\rm{H}}+ {\bf{I}}_N\nonumber \\
&-(
{\bf{P}}^{\rm{H}}{\bf{\bar
H}}_{sr}^{\rm{H}}{\bf{F}}^{\rm{H}}{\bf{\bar
H}}_{rd}^{\rm{H}}{\bf{G}}^{\rm{H}})- (
{\bf{G}}{\bf{\bar H}}_{rd}{\bf{F}}{\bf{\bar H}}_{sr}{\bf{P}}
),
\end{align}where matrices ${\bf{R}}_{\bf{x}}$ and ${\bf{K}}_2$ are defined as\begin{align}
\label{R_x}
{\bf{R}}_{\bf{x}}&\triangleq \mathbb{E}\{{\bf{x}}{\bf{x}}^{\rm{H}}\}={\bf{\bar H}}_{sr}{\bf{P}}{\bf{P}}^{\rm{H}}{\bf{\bar H}}_{sr}^{\rm{H}}+{\bf{K}}_1 \nonumber \\{\bf{K}}_1& \triangleq {\rm{Tr}}({\bf{P}}{\bf{P}}^{\rm{H}}{\boldsymbol
\Psi}_{sr}){\boldsymbol \Sigma}_{sr}+{\bf{R}}_{n_1}\nonumber \\
{\bf{K}}_2&\triangleq {\rm{Tr}}({\bf{F}}{\bf{R}}_{\bf{x}} {\bf{F}}^{\rm{H}}
{\boldsymbol\Psi} _{rd}
   ){\boldsymbol\Sigma} _{rd}+{\bf{R}}_{n_2}.
\end{align}It is obvious that ${\bf{R}}_{\bf{x}}$ is the covariance matrix of the received signal at the relay. Using linear Bayesian theory, the LMMSE equalizer at the destination equals to
\begin{align}
\label{G} & {\bf{G}}=({\bf{\bar H}}_{rd}{\bf{
F}}{\bf{\bar
H}}_{sr}{\bf{P}})^{\rm{H}}({\bf{\bar H}}_{rd}{\bf{
F}}{\bf{R}}_{\bf{x}}{\bf{ F}}^{\rm{H}}{\bf{\bar
H}}_{rd}^{\rm{H}}+{\bf{K}}_2)^{-1},
\end{align}based on which the MSE matrix in (\ref{MSE_final}) is rewritten as
\begin{align}
\label{MSE_Matrix_Orig}
&{\boldsymbol {\Phi}}_{\rm{MSE}}({\bf{
F}},{\bf{P}})\nonumber \\
  =&{\bf{I}}-({\bf{\bar
H}}_{rd}{\bf{
F}}{\bf{\bar
H}}_{sr}{\bf{P}})^{\rm{H}}({\bf{\bar H}}_{rd}{\bf{
F}}{\bf{R}}_{\bf{x}}{\bf{ F}}^{\rm{H}} {\bf{\bar
H}}_{rd}^{\rm{H}}+{\bf{K}}_2)^{-1}\nonumber \\
&\times ({\bf{\bar H}}_{rd}{\bf{
F}}{\bf{\bar H}}_{sr}{\bf{P}})
\end{align}

Capacity is one of the most important and widely used performance metrics for transceiver designs. Given the received pilots in channel estimation as ${\bf{y}}_1$ and ${\bf{y}}_2$, the channel capacity is denoted as  $\textsl{I}({\bf{s}};{\bf{y}}|{\bf{y}}_1,{\bf{y}}_2)$, which is the conditional mutual information based on known ${\bf{y}}_1$ and ${\bf{y}}_2$ \cite{Ding10}. To the best of our knowledge, the exact capacity for MIMO channels with estimation errors at both ends is largely open even for point-to-point MIMO systems \cite{Ding10}. To proceed, a common logic is to derive and use bounds i.e., lower bound or upper bound. Since we aim to maximize channel capacity, lower bound is more meaningful than upper bound. In Appendix~\ref{Appedix:5} it has been proved that
\begin{align}
 -{\rm{log}}|{\boldsymbol {\Phi}}_{\rm{MSE}}({\bf{
F}},{\bf{P}}) | \le \textsl{I}({\bf{s}};{\bf{y}}|{\bf{y}}_1,{\bf{y}}_2).
\end{align} This is a widely-established lower bound \cite{Ding10} and becomes tighter as estimation errors are smaller.

As a result, the robust transceiver design for maximizing mutual information is formulated as
\begin{align}
\label{prob:opt}
& \min_{{\bf{F}},{\bf{P}}} \ \ {\rm{log}}|{\boldsymbol {\Phi}}_{\rm{MSE}}({\bf{
F}},{\bf{P}}) | \nonumber \\
& \   {\rm{s.t.}} \ \ \ {\rm{Tr}}({\bf{F}}{\bf{R}}_{\bf{x}}{\bf{F}}^{\rm{H}}) \le P_r \ \ {\rm{Tr}}({\bf{P}}{\bf{P}}) \le P_s.
\end{align} Based on the definition of ${\bf{R}}_{\bf{x}}$ in (\ref{R_x}), ${\bf{R}}_{{\bf{x}}}$ is a function of ${\bf{P}}$. In order to simplify the analysis, we define a new variable\begin{align}
\label{Tilde_F}
{\bf{\tilde F}}\triangleq
{\bf{F}}{\bf{K}}_1^{1/2}(\underbrace{{\bf{K}}_1^{-1/2}{\bf{\bar
H}}_{sr}{\bf{P}}{\bf{P}}^{\rm{H}}{\bf{\bar
H}}_{sr}^{\rm{H}}{\bf{K}}_1^{-1/2}+{\bf{I}}}_{\triangleq {\boldsymbol \Pi}_{\bf{P}}})^{1/2},
\end{align}based on which ${\bf{F}}{\bf{R}}_{\bf{x}}{\bf{F}}^{\rm{H}}={\bf{\tilde F}}{\bf{\tilde F}}^{\rm{H}}$ and the two constraints involved in (\ref{prob:opt}) become independent. Meanwhile, the MSE matrix in (\ref{MSE_Matrix_Orig}) is correspondingly
rewritten as
\begin{align}
\label{MSE_MATRIX}
{\boldsymbol { \Phi}}_{\rm{MSE}}({\bf{\tilde
F}},{\bf{P}}) & ={\bf{I}}-({\bf{\bar
H}}_{rd}{\bf{\tilde
F}}{\boldsymbol \Pi}_{\bf{P}}^{-1/2}{\bf{K}}_1^{-1/2}{\bf{\bar
H}}_{sr}{\bf{P}})^{\rm{H}}({\bf{\bar H}}_{rd}{\bf{\tilde
F}}{\bf{\tilde F}}^{\rm{H}} \nonumber \\
 \times{\bf{\bar
H}}_{rd}^{\rm{H}}&+{\bf{K}}_2)^{-1}({\bf{\bar H}}_{rd}{\bf{\tilde
F}}{\boldsymbol \Pi}_{\bf{P}}^{-1/2}{\bf{K}}_1^{-1/2}{\bf{\bar H}}_{sr}{\bf{P}}).
\end{align}Finally, the optimization problem for the robust design becomes
\begin{align}
\label{opt:capacity}
& \min_{{\bf{\tilde F}},{\bf{P}}} \ \ {\rm{log}}|{\boldsymbol {\Phi}}_{\rm{MSE}}({\bf{\tilde
F}},{\bf{P}}) | \nonumber \\
& \   {\rm{s.t.}} \ \ \ {\rm{Tr}}({\bf{\tilde F}}{\bf{\tilde F}}^{\rm{H}}) \le P_r \ \ {\rm{Tr}}({\bf{P}}{\bf{P}}) \le P_s.
\end{align}In the following, the optimal solutions of (\ref{opt:capacity}) will be discussed in details.

\section{Optimal Solutions}

In our work, we investigate the optimization problem (\ref{opt:capacity}) from matrix-monotone function viewpoint. The idea of utilizing the properties of matrix-monotone functions to design MIMO transceivers has been address in \cite{Jorswieck07}. In this paper, we extend this idea to robust transceiver designs for a AF MIMO relaying system.

\noindent \textbf{Definition 1:} A matrix-monotone function is defined as ${\boldsymbol g}(\bullet)$ which maps a matrix variable from a subsect of positive semi-definite matrices
 to a real number. If ${\boldsymbol g}(\bullet)$ is a monotonically decreasing matrix-monotone function on positive semi-definite matrices, it satisfies
\begin{align}
{\boldsymbol{A}} \succeq {\boldsymbol B}\succeq {\bf{0}} \rightarrow {\boldsymbol g}({\boldsymbol{A}}) \le {\boldsymbol g}({\boldsymbol{B}}).
\end{align}On the other hand, when ${\boldsymbol g}(\bullet)$ a monotonically increasing matrix-monotone, it means $-{\boldsymbol g}(\bullet)$ is a monotonically decreasing matrix-monotone function \cite{Jorswieck07}.

In the following we focus our attention on a kind of optimization problems with  a decreasing matrix-monotone function as objective, which is formulated as
\begin{align}
\label{Opt_X}
& \min_{{\bf{X}}} \ \ \  {\boldsymbol g}\left({\bf{X}}^{\rm{H}}{\bf{H}}^{\rm{H}}{\bf{H}}{\bf{X}}\frac{1}{\eta_x}\right) \nonumber \\
& \ {\rm{s.t.}} \ \ \ \ {\rm{Tr}}({\bf{X}}{\bf{X}}^{\rm{H}}) \le P \ \  \eta_x={\rm{Tr}}({\bf{X}}{\bf{X}}^{\rm{H}}{\boldsymbol \Psi})\alpha+{\sigma}_{n}^2.
\end{align} Solving the optimization problem (\ref{Opt_X}), two important solutions are derived and are the basis for the following derivations.

\noindent \textbf{\underline{\textsl{Conclusion 1:}}} The optimal solution of (\ref{Opt_X}) satisfies
\begin{align}
{\rm{Tr}}({\bf{X}}{\bf{X}}^{\rm{H}})={\rm{Tr}}[{\bf{X}}
{\bf{X}}^{\rm{H}}(\alpha P{\boldsymbol \Psi}+\sigma_n^2{\bf{I}})]/\eta_x=P.
\end{align}Defining the unitary matrix ${\bf{V}}_{\bf{H}}$ and rectangular diagonal matrix ${\boldsymbol \Lambda}_{\bf{H}}$ based on the following singular value decomposition
\begin{align}
{\bf{H}}(\alpha P{\boldsymbol \Psi}+{\sigma}_{n}^2{\bf{I}})^{-1/2}={\bf{U}}_{\bf{H}}{\boldsymbol \Lambda}_{\bf{H}}{\bf{V}}_{\bf{H}}^{\rm{H}} \ \ \text{with} \ \  {\boldsymbol \Lambda}_{\bf{H}} \searrow,
\end{align}the optimal solution of the optimization problem (\ref{Opt_X}) has following structure
\begin{align}
\label{conslusion_1}
&{\bf{X}}_{\rm{opt}}=\sqrt{\eta_x}(\alpha P{\boldsymbol
\Psi}+\sigma_{n}^2{\bf{I}})^{-1/2}{\bf{V}}_{\bf{H}}{\boldsymbol \Lambda}_{\bf{X}}{\bf{U}}_{\boldsymbol{\Xi}}^{\rm{H}} \nonumber \\
&\text{with}  \ \ {\boldsymbol \Lambda}_{\bf{X}}^{\rm{T}}{\boldsymbol \Lambda}_{\bf{H}}^{\rm{T}}{\boldsymbol \Lambda}_{\bf{H}}{\boldsymbol \Lambda}_{\bf{X}} \searrow, \nonumber \\
&\text{and}\ \ \eta_x={\sigma_{n}^2}/\{1-\alpha {\rm{Tr}}[{\bf{V}}_{\bf{H}}^{\rm{H}}(\alpha P{\boldsymbol
\Psi}+\sigma_{n}^2{\bf{I}})^{-1/2}\nonumber \\
&\times{\boldsymbol
\Psi}(\alpha P{\boldsymbol
\Psi}+\sigma_{n}^2{\bf{I}})^{-1/2}
{\bf{V}}_{\bf{H}}{\boldsymbol{\Lambda}}_{{\bf{X}}}{\boldsymbol{\Lambda}}_{{\bf{X}}}^{\rm{T}}]\},
\end{align}where ${\bf{U}}_{\boldsymbol{\Xi}}$ is an unitary matrix and ${\boldsymbol \Lambda}_{\bf{X}}$ is a rectangular diagonal matrix with real diagonal elements.

\textsl{Proof:} See Appendix~\ref{Appedix:3}. $\blacksquare$

\noindent \textbf{\underline{\textsl{Conclusion 2:}}}
For a complex matrix ${\bf{A}}$ and a positive define matrix ${\bf{N}}$, based on following eigen-decomposition
\begin{align}
&  {\bf{A}}{\bf{N}}^{-1}{\bf{A}}^{\rm{H}}={\bf{U}}_{\bf{ANA}}{\boldsymbol \Lambda}_{\bf{ANA}}
{\bf{V}}_{\bf{ANA}}^{\rm{H}}\ \ {\text{with}} \ \ {\boldsymbol \Lambda}_{\bf{ANA}} \searrow
\end{align}when the objective function is
\begin{align}
\label{objective}
{\boldsymbol g}\left({\bf{X}}^{\rm{H}}{\bf{H}}^{\rm{H}}{\bf{H}}{\bf{X}}/\eta_x\right)
=
{\rm{log}}|{\bf{A}}^{\rm{H}}({\bf{X}}^{\rm{H}}{\bf{H}}^{\rm{H}}{\bf{H}}{\bf{X}}/\eta_x
+{\bf{I}})^{-1}{\bf{A}}+{\bf{N}}|\end{align} the unitary matrix ${\bf{U}}_{\boldsymbol{\Xi}}$ in (\ref{conslusion_1}) equals to
\begin{align}
&{\bf{U}}_{\boldsymbol{\Xi}}= {\bf{U}}_{\bf{ANA}}.
\end{align}
Defining $N_X=\min\{{\rm{Rank}}$$({\bf{H}}^{\rm{H}}{\bf{H}}), {\rm{Rank}}({\bf{A}}{\bf{A}}^{\rm{H}}) \}$, ${\boldsymbol \Lambda}_{\bf{x}}$ in (\ref{conslusion_1}) has the structure of
\begin{align}
{\boldsymbol \Lambda}_{\bf{x}}=
\left[ {\begin{array}{*{20}c}
   {{\boldsymbol {\tilde \Lambda}}_{\bf{x}}} & {{\bf{0}}}  \\
   {{\bf{0}}} & {{\bf{0}}}  \\
\end{array}} \right],
\end{align}where ${\boldsymbol {\tilde \Lambda}}_{\bf{x}}$ is a $N_X \times N_X$ diagonal matrix.

\textsl{Proof:} See Appendix~\ref{Appedix:4}. $\blacksquare$

%


\subsection{The structure of optimal ${\bf{\tilde F}}$}
In this section the structure of optimal ${\bf{\tilde F}}$ is derived.
Based on the matrix inversion lemma, the MSE matrix in (\ref{MSE_MATRIX}) can be rewritten as
\begin{align}
&{\boldsymbol {\Phi}}_{\rm{MSE}}({\bf{\tilde
F}},{\bf{P}})\nonumber \\=&({\boldsymbol \Pi}_{\bf{P}}^{-1/2}{\bf{K}}_1^{-1/2}{\bf{\bar
H}}_{sr}{\bf{P}})^{\rm{H}}({\bf{\tilde F}}^{\rm{H}}{\bf{\bar H}}_{rd}^{\rm{H}}{\bf{K}}_2^{-1}{\bf{\bar H}}_{rd}{\bf{\tilde F}}+{\bf{I}})^{-1}\nonumber \\
& \times (\underbrace{{\boldsymbol \Pi}_{\bf{P}}^{-1/2}{\bf{K}}_1^{-1/2}{\bf{\bar
H}}_{sr}{\bf{P}}}_{\triangleq {\bf{A}}_{\bf{P}}})+\underbrace{({\bf{P}}^{\rm{H}}{\bf{\bar
H}}_{sr}^{\rm{H}}{\bf{K}}_1^{-1}{\bf{\bar
H}}_{sr}{\bf{P}}+{\bf{I}})^{-1}}_{\triangleq {\bf{N}}_{\bf{P}}} \nonumber
\end{align} based on which for any given ${\bf{P}}$ the optimization problem with respect to ${\bf{\tilde F}}$ becomes as
\begin{align}
 & \min_{{\bf{\tilde F}}} \ \  {\rm{log}}|{\bf{A}}_{\bf{P}}^{\rm{H}}
({\bf{\tilde F}}^{\rm{H}}{\bf{\bar H}}_{rd}^{\rm{H}}{\bf{K}}_2^{-1}{\bf{\bar H}}_{rd}{\bf{\tilde F}}+{\bf{I}})^{-1}{\bf{A}}_{\bf{P}}+{\bf{N}}_{\bf{P}}| \nonumber \\
& \   {\rm{s.t.}} \ \ \ {\rm{Tr}}({\bf{\tilde F}}{\bf{\tilde F}}^{\rm{H}}) \le P_r \ \ {\bf{K}}_2={\rm{Tr}}({\bf{\tilde F}}{\bf{\tilde F}}^{\rm{H}}{\boldsymbol \Psi}_{rd}){\boldsymbol \Sigma}_{rd}+\sigma_{n_2}^2{\bf{I}} .
\end{align}

Defining unitary matrices ${{\bf{U}}_{1}}$ and ${{\bf{V}}_{1}}$ based on the following singular value decomposition
\begin{align}
\label{U_1}
{\bf{K}}_1^{-1/2}{\bf{\bar
H}}_{sr}{\bf{P}}={\bf{U}}_{1}{\boldsymbol \Lambda}_{1}{\bf{V}}_{1}^{\rm{H}}  \ {\text{with}} \ {\boldsymbol \Lambda}_{1} \searrow,
\end{align} we have the following eigen-decomposition
\begin{align}
&{\bf{A}}_{\bf{P}}{\bf{N}}_{\bf{P}}^{-1}{\bf{A}}_{\bf{P}}^{\rm{H}}
={\bf{U}}_{1}
{\boldsymbol { \Lambda}}_{1}{\boldsymbol { \Lambda}}_{1}^{\rm{T}}{\bf{U}}_{1}^{\rm{H}}  \ {\text{with }} \ {\boldsymbol { \Lambda}}_{1}{\boldsymbol{\Lambda}}_{1}^{\rm{T}} \searrow.
\end{align}Together with the following singular value decomposition
\begin{align}
&{\bf{\bar H}}_{rd}(\alpha_2P_r{\boldsymbol \Psi}_{rd}+{\sigma_{n_2}^2})^{-1/2}={\bf{U}}_{rd}{\boldsymbol \Lambda}_{rd}{\bf{V}}_{rd}^{\rm{H}} \ {\text{with }} \ {\boldsymbol \Lambda}_{rd} \searrow, \nonumber
\end{align}and based on \textbf{Conclusions 1} and \textbf{2}\footnote{Notice that ${\boldsymbol \Sigma}_{rd}=\alpha_2{\bf{I}}.$}, the optimal ${\bf{\tilde F}}$ has the following structure
\begin{align}
{\bf{\tilde F}}&=\sqrt{\eta_f}(\alpha_2P_{r}{\boldsymbol
\Psi}_{rd}+\sigma_{n_2}^2{\bf{I}})^{-1/2}{\bf{V}}_{rd,N}
{\boldsymbol{\tilde \Lambda}}_{\bf{\tilde F}}
{\bf{U}}_{1,N}^{\rm{H}} \nonumber \\
& {\text{with}} \ \ {\boldsymbol{\tilde \Lambda}}_{\bf{\tilde F}}{\boldsymbol{\tilde \Lambda}}_{rd}^2{\boldsymbol{\tilde \Lambda}}_{\bf{\tilde F}} \searrow \nonumber \\
& {\text{and}} \ \ \eta_f={\sigma_{n_2}^2}/\{1-\alpha_2{\rm{Tr}}[{\bf{V}}_{rd,N}^{\rm{H}}(\alpha_2P_{r}{\boldsymbol
\Psi}_{rd}+\sigma_{n_2}^2{\bf{I}})^{-1/2}\nonumber \\
& \times {\boldsymbol
\Psi}_{rd}(\alpha_2P_r{\boldsymbol
\Psi}_{rd}+\sigma_{n_2}^2{\bf{I}})^{-1/2}
{\bf{V}}_{rd,N}{\boldsymbol{\tilde \Lambda}}_{{\bf{\tilde F}}}^2]\},
\end{align}where ${\boldsymbol{\tilde \Lambda}}_{\bf{\tilde F}}$ is a $N\times N$ diagonal matrix.  Meanwhile, for the optimal ${\bf{\tilde F}}$ the following constraint is fulfilled
\begin{align}
\label{cons_F}
{\rm{Tr}}({\bf{\tilde F}}{\bf{\tilde F}}^{\rm{H}})={\rm{Tr}}[{\bf{\tilde F}}{\bf{\tilde F}}^{\rm{H}}(\alpha_2P_r{\boldsymbol \Psi}_{rd}+\sigma_{n_2}^2{\bf{I}})]/\eta_f=P_r.
\end{align}

\subsection{The structure of optimal ${\bf{P}}$}
In the following, it will be proved that given the structure of ${\bf{\tilde F}}$ the optimization problem for ${\bf{P}}$ is the same as that for ${\bf{\tilde F}}$. Using the optimal structure of ${\bf{\tilde F}}$, we have
\[
({\bf{\tilde F}}^{\rm{H}}{\bf{\bar H}}_{rd}^{\rm{H}}{\bf{K}}_2^{-1}{\bf{\bar H}}_{rd}{\bf{\tilde F}}+{\bf{I}})^{-1}
=({\bf{U}}_{1,N}{\boldsymbol {\tilde \Lambda}}_{\bf{\tilde F}}{\boldsymbol {\tilde \Lambda}}_{rd}^2{\boldsymbol {\tilde \Lambda}}_{\bf{\tilde F}}{\bf{U}}_{1,N}^{\rm{H}}+{\bf{I}})^{-1}.\]
Using the following substitution
\begin{align}
{\boldsymbol \Lambda}_2\triangleq ({\boldsymbol {\tilde \Lambda}}_{\bf{\tilde F}}{\boldsymbol {\tilde \Lambda}}_{rd}^2{\boldsymbol {\tilde \Lambda}}_{\bf{\tilde F}}+{\bf{I}})^{-1}\nearrow,
\end{align}and the matrix inversion lemma again, the MSE matrix can be reformulated as
\begin{small}\begin{align}
{\boldsymbol {\Phi}}_{\rm{MSE}}({\bf{\tilde
F}},{\bf{P}})
=&{\bf{V}}_{1}({\bf{I}}-{\boldsymbol \Lambda}_2)^{1/2}{\bf{V}}_{1}^{\rm{H}}{({\bf{P}}^{\rm{H}}{\bf{\bar
H}}_{sr}^{\rm{H}}{\bf{K}}_1^{-1}{\bf{\bar
H}}_{sr}{\bf{P}}+{\bf{I}})^{-1}} \nonumber \\
& \times \underbrace{{\bf{V}}_{1}({\bf{I}}-{\boldsymbol \Lambda}_2)^{1/2}{\bf{V}}_{1}^{\rm{H}}}_{\triangleq {\bf{A}}_{\bf{\tilde F}}}+\underbrace{{\bf{V}}_{1}{\boldsymbol \Lambda}_2{\bf{V}}_{1}^{\rm{H}}}_{\triangleq {\bf{N}}_{\bf{\tilde F}}}.
\end{align}\end{small}Therefore, the optimization problem with respective to ${\bf{P}}$ is equivalent to
\begin{align}
& \min_{{\bf{P}}} \ \  {\rm{log}}|{\bf{A}}_{\bf{\tilde F}}^{\rm{H}}
({\bf{P}}^{\rm{H}}{\bf{\bar H}}_{sr}^{\rm{H}}{\bf{K}}_1^{-1}{\bf{\bar H}}_{sr}{\bf{P}}+{\bf{I}})^{-1}{\bf{A}}_{\bf{\tilde F}}+{\bf{N}}_{\bf{\tilde F}}| \nonumber \\
& \   {\rm{s.t.}} \ \  {\rm{Tr}}({\bf{P}}{\bf{P}}^{\rm{H}}) \le P_s,  \ {\bf{K}}_1={\rm{Tr}}({\bf{P}}{\bf{P}}^{\rm{H}}{\boldsymbol \Psi}_{sr}){\boldsymbol \Sigma}_{sr}+\sigma_{n_1}^2{\bf{I}}.
\end{align}Based on the definitions of ${\bf{A}}_{\bf{\tilde F}}$ and ${\bf{N}}_{\bf{\tilde F}}$, it can be derived that ${\bf{A}}_{\bf{\tilde F}}{\bf{N}}_{\bf{\tilde F}}^{-1}{\bf{A}}_{\bf{\tilde F}}^{\rm{H}}$ has the following eigen-decomposition
\begin{align}
{\bf{A}}_{\bf{\tilde F}}{\bf{N}}_{\bf{\tilde F}}^{-1}{\bf{A}}_{\bf{\tilde F}}^{\rm{H}}={\bf{V}}_{1}{\boldsymbol {\tilde \Lambda}}_{\bf{\tilde F}}{\boldsymbol {\tilde \Lambda}}_{rd}^2{\boldsymbol {\tilde \Lambda}}_{\bf{\tilde F}}{\bf{V}}_{1}^{\rm{H}}.
\end{align} Together with
following singular value decomposition,
\begin{align}
& {\bf{\bar{H}}}_{sr}(\alpha_1P_{s}{\boldsymbol
\Psi}_{sr}+\sigma_{n_1}^2{\bf{I}})^{-1/2}={\bf{U}}_{sr}{\boldsymbol
\Lambda}_{sr}{\bf{V}}_{sr}^{\rm{H}}
\end{align}and using \textbf{Conclusions 1} and \textbf{2}\footnote{Notice that ${\boldsymbol \Sigma}_{sr}=\alpha_1{\bf{I}}$.}, the optimal ${\bf{P}}$ has the following structure
\begin{align}
\label{Structure_P_capacity}
&{\bf{P}}=\sqrt{\eta_p}(\alpha_1P_{s}{\boldsymbol
\Psi}_{sr}+\sigma_{n_1}^2{\bf{I}})^{-1/2}{\bf{V}}_{sr,N}
{\boldsymbol{\tilde \Lambda}}_{\bf{P}}
{\bf{V}}_{1}^{\rm{H}} \nonumber \\
& {\text{with}} \ \ {\boldsymbol{\tilde \Lambda}}_{\bf{P}}{\boldsymbol{\tilde \Lambda}}_{sr}^2{\boldsymbol{\tilde \Lambda}}_{\bf{P}} \searrow \nonumber \\
& {\text{and}} \ \ \eta_p={\sigma_{n_1}^2}/\{1-\alpha_1{\rm{Tr}}[{\bf{V}}_{sr,N}^{\rm{H}}(\alpha_1P_{s}{\boldsymbol
\Psi}_{sr}+\sigma_{n_1}^2{\bf{I}})^{-1/2}\nonumber \\
& \times {\boldsymbol
\Psi}_{sr}(\alpha_1P_s{\boldsymbol
\Psi}_{sr}+\sigma_{n_1}^2{\bf{I}})^{-1/2}
{\bf{V}}_{sr,N}{\boldsymbol{\tilde \Lambda}}_{{\bf{P}}}^2]\},
\end{align}where ${\boldsymbol{\tilde \Lambda}}_{\bf{P}}$ a $N\times N$ diagonal matrix. Considering that as there are no constraints on ${\bf{V}}_{1}$,  ${\bf{V}}_{1}$ can be an arbitrary $N \times N$ unitary matrix. Based on \textbf{Conclusion 1}, it can be concluded that the optimal ${\bf{P}}$ satisfies
\begin{align}
\label{cons_P}
{\rm{Tr}}({\bf{P}}{\bf{P}}^{\rm{H}})={\rm{Tr}}[{\bf{P}}{\bf{P}}^{\rm{H}}(\alpha_1P_s{\boldsymbol \Psi}_{sr}+\sigma_{n_1}^2{\bf{I}})]/\eta_p=P_s.
\end{align}

Substituting (\ref{Structure_P_capacity}) into (\ref{U_1}), it can be derived that ${\bf{U}}_{1,N}={\bf{U}}_{sr,N}$ and then the optimal structure of ${\bf{\tilde F}}$ is
\begin{align}
\label{Structure_F_capacity}
{\bf{\tilde F}}&=\sqrt{\eta_f}(\alpha_2P_{r}{\boldsymbol
\Psi}_{rd}+\sigma_{n_2}^2{\bf{I}})^{-1/2}{\bf{V}}_{rd,N}
{\boldsymbol{\tilde \Lambda}}_{\bf{\tilde F}}
{\bf{U}}_{sr,N}^{\rm{H}} .
\end{align}
Based on the optimal structure given by (\ref{Structure_P_capacity}) and (\ref{Structure_F_capacity}) and with regard to the fact that $\eta_p$ and $\eta_f$ are determined by ${\boldsymbol{\tilde \Lambda}}_{\bf{P}}$ and ${\boldsymbol{\tilde \Lambda}}_{\bf{\tilde F}}$, respectively, the left unknown variables are only ${\boldsymbol{\tilde \Lambda}}_{\bf{P}}$ and ${\boldsymbol{\tilde \Lambda}}_{\bf{\tilde F}}$.

\subsection{Proposed Solutions for ${\boldsymbol \Lambda}_{\bf{\tilde F}}$ and ${\boldsymbol \Lambda}_{\bf{\tilde P}}$}

Based on (\ref{cons_F}) and (\ref{cons_P}), the optimization problem (\ref{opt:capacity}) also equals to
\begin{align}
\label{opt:capacity_a}
& \min_{{\bf{\tilde F}},{\bf{P}}} \ \ {\rm{log}}|{\boldsymbol {\Phi}}_{\rm{MSE}}({\bf{\tilde
F}},{\bf{P}}) | \nonumber \\
& \  {\rm{s.t.}} \ \ \ {\rm{Tr}}[{\bf{\tilde F}}{\bf{\tilde F}}^{\rm{H}}(\alpha_2P_r{\boldsymbol \Psi}_{rd}+\sigma_{n_2}^2{\bf{I}})]/\eta_f= P_r \nonumber \\  & \ \ \ \ \ \ \ \ {\rm{Tr}}[{\bf{P}}{\bf{P}}^{\rm{H}}(\alpha_1P_s{\boldsymbol \Psi}_{sr}+\sigma_{n_1}^2{\bf{I}})]/\eta_p=P_s.
\end{align}Furthermore, with the following diagonal matrices
\begin{align}
& {\boldsymbol { \tilde \Lambda}}_{sr}={\rm{diag}}\{\lambda_{sr,i}\} \ \
{\boldsymbol {\tilde  \Lambda}}_{rd}={\rm{diag}}\{\lambda_{rd,i}\} \nonumber \\
&
{\boldsymbol {\tilde \Lambda}}_{{\bf{\tilde F}}}={\rm{diag}}\{f_{i}\} \ \
{\boldsymbol {\tilde \Lambda}}_{{\bf{ P}}}={\rm{diag}}\{p_{i}\}
\end{align}and substituting (\ref{Structure_P_capacity}) and (\ref{Structure_F_capacity}) into (\ref{opt:capacity_a}), the optimization problem (\ref{opt:capacity_a}) can be rewritten as
\begin{align}
\label{opt:capacity_b}
& \min_{f_i,p_i} \ \ \ \sum_{i=1}^{N}{\rm{log}} \frac{f_i^2\lambda_{rd,i}^2+p_i^2\lambda_{sr,i}^2+1}{(p_i^2\lambda_{sr,i}^2+1)
(f_i^2\lambda_{rd,i}^2+1)} \nonumber \\
& \ {\rm{s.t.}} \ \ \ \ \sum_i f_i^2 =P_r \ \ \sum_i p_i^2=P_s.
\end{align}With respective to the fact the problem (\ref{opt:capacity_b}) is inherently non-convex and difficult to solve, an iterative water-filling solution is proposed in this paper. When $p_i$'s are fixed, $f_i$'s can be computed as
\begin{small}
\begin{align}
&f_i^2=\left( \frac{-p_i^2\lambda_{sr,i}^2+
\sqrt{(p_i^2\lambda_{sr,i}^2)^2+\frac{4p_i^2\lambda_{sr,i}^2\lambda_{rd,i}^2}{\mu_f}}}{2\lambda_{rd,i}^2}-\frac{1}{\lambda_{rd,i}^2} \right)^{+}
\end{align}\end{small}where $\mu_f\ge 0$ is the Lagrange multiplier which makes $\sum_i f_i^2 =P_r$. On the other hand, when $f_i$'s are fixed $p_i$'s can be computed as
\begin{small}
\begin{align}
&p_i^2=\left( \frac{-f_i^2\lambda_{rd,i}^2+
\sqrt{(f_i^2\lambda_{rd,i}^2)^2+\frac{4f_i^2\lambda_{rd,i}^2\lambda_{sr,i}^2}{\mu_p}}}{2\lambda_{sr,i}^2}-\frac{1}{\lambda_{sr,i}^2} \right)^{+}
\end{align}\end{small}where $\mu_p\ge0$ is the Lagrange multiplier which makes $\sum_i p_i^2 =P_s$.

\noindent {\textbf{Special cases}}: Several existing algorithms can be considered as special cases of our proposed solution.

\noindent $\bullet$ When CSI is perfectly known and ${\bf{P}}={\bf{I}}$, the proposed solution for ${\bf{F}}$ reduces to that in \cite{Tang07}.

\noindent $\bullet$ When CSI is perfectly known, the proposed solution for ${\bf{P}}$ and ${\bf{F}}$ reduces to that given in \cite{Rong09}.

\noindent $\bullet$ When the second hop channel is an identity matrix and noiseless, the proposed solution for source precoder design reduces to that given in \cite{Ding10}.

\section{Simulation Results and Discussions}

In this section, simulation results are presented to assess the
performance of the proposed algorithm and for the purpose of
comparison, the algorithm based on the estimated channel only
(without taking the channel errors into account) \cite{Rong09}. In the following, we consider an AF MIMO relay
system where the source, relay and destination are equipped with
same number of antennas, i.e., $N_S=M_R=N_R=M_D=4$. The
channels ${\bf{H}}_{sr}$ and ${\bf{H}}_{rd}$ are randomly generated
according to i.i.d. Gaussian distribution.

To estimate the channels, a practical LMMSE estimation algorithm is adopted \cite{Xing1012}.
For the training sequence ${\bf{D}}$, a famous exponential correlation matrix is used to describe the correlation matrix of ${\bf{D}}$, i.e., ${\bf{D}}{\bf{D}}^{\rm{H}}\propto {\bf{R}}_{\rho}$ where $[{\bf{R}}_{\rho}]_{ij}=\rho^{|i-j|}$. As a result, ${\boldsymbol \Sigma}_{sr}={\boldsymbol \Sigma}_{rd}={\bf{I}}$ and ${\boldsymbol \Psi}_{sr}={\boldsymbol \Psi}_{rd}=({\bf{I}}_4+{\rm{SNR}}_{\rm{EST}}{\bf{R}}_{\alpha})^{-1}$ where ${\rm{SNR}}_{\rm{EST}}$ is the signal-to-noise ratio (SNR) in channel estimation process \cite{Xing1012}\footnote{The detailed derivation is given in \cite{Xing1012}}.

In the simulation,\begin{figure}[!ht] \centering
\includegraphics[width=.4\textwidth]{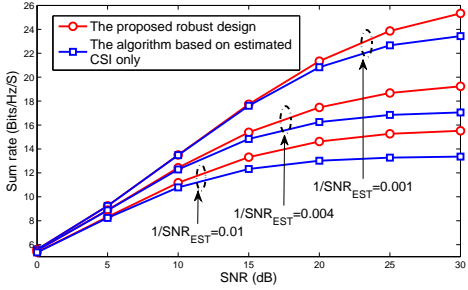}
\caption{Sum rates of different
algorithms when $\rho=0.5$.}\label{fig:1}
\end{figure} for data transmission process the SNR at relay is defined as
$P_s/{\sigma_{1}}^2$, and the SNR at destination is
defined as $P_r/\sigma_2^2$. For simplicity, it is also assumed that $P_s/{\sigma_{1}}^2=P_r/\sigma_2^2$. Each point in the
following figure is an average of 10000 independent channel
realizations.

Fig.~\ref{fig:1} shows the sum rates of different algorithms including
the proposed robust design and its counterpart based on estimated CSI only when $\rho=0.5$. It can be seen that
the performance of the proposed robust design is always better than that of the design based estimated on CSI only. Furthermore, as the channel estimation SNR decreases the performance gain of the robust design becomes larger.
\section{Conclusions}

Robust mutual information maximization transceiver design for
dual-hop AF MIMO relay systems was investigated. With Gaussian distributed channel errors the precoder at the source and forwarding matrix at
the relay were jointly designed. The structures of the optimal solutions were derived first, which differentiates our work from the existing works. For the unknown diagonal matrices, a well-known iterative water-filling solution was proposed. The simulation result demonstrated the performance advantage of our robust design.

\appendices

\section{Lower Bound of Capacity}
\label{Appedix:5}
Denoting ${\bf{y}}_1$ and ${\bf{y}}_2$ as the received pilots in the separate dual hop channel estimations, the capacity between the source and destination equals to
\begin{align}
\label{capacity_equ}
\textsl{I}({\bf{s}};{\bf{y}}|{\bf{y}}_1,{\bf{y}}_2)=\underbrace{\textsl{H}({\bf{s}}|{\bf{y}}_1,{\bf{y}}_2)}_{=\textsl{H}({\bf{s}})}-\textsl{H}({\bf{s}}|{\bf{y}},{\bf{y}}_1,{\bf{y}}_2)
\end{align}where $\textsl{H}({\bf{s}}|{\bf{y}}_1,{\bf{y}}_2)$ denotes the conditional entropy of ${\bf{s}}$ when ${\bf{y}}_1$ and ${\bf{y}}_2$ are known and $\textsl{H}({\bf{s}}|{\bf{y}},{\bf{y}}_1,{\bf{y}}_2)$ is the conditional entropy of ${\bf{s}}$ when ${\bf{y}}$, ${\bf{y}}_1$ and ${\bf{y}}_2$ are known. As ${\bf{y}}_1$ and ${\bf{y}}_2$ are independent with ${\bf{s}}$, $\textsl{H}({\bf{s}}|{\bf{y}}_1,{\bf{y}}_2)=\textsl{H}({\bf{s}})$.
The second term on the right hand side of (\ref{capacity_equ}) will be discussed in the following.

Denoting ${\bf{\bar s}}={\mathbb{E}}_{|{\bf{y}},{\bf{y}}_1,{\bf{y}}_2}\{s\}$ as the conditional mean and based on the definition of covariance matrix, the conditional covariance matrix satisfies
\begin{align}
\label{covariance_matrix}
{\boldsymbol{\rm{Cov}}}({\bf{s}}|{\bf{y}},{\bf{y}}_1,{\bf{y}}_2)&=\mathbb{E}\{({\bf{s}}-{\bf{\bar s}})({\bf{s}}-{\bf{\bar s}})^{\rm{H}}\} \nonumber \\
&\preceq  \mathbb{E}\{({\bf{s}}-{\bf{\bar s}})({\bf{s}}-{\bf{\bar s}})^{\rm{H}}\}+
\mathbb{E}\{({\bf{\bar s}}-{\bf{\hat s}})({\bf{\bar s}}-{\bf{\hat s}})^{\rm{H}}\}
\end{align}where ${\bf{\hat s}}$ is an arbitrary estimate of ${\bf{s}}$ including LMMSE estimate. It should be pointed out that as CSI is not perfectly known, ${\bf{\hat s}}\not={\bf{\bar s}}$. Notice that ${\bf{\bar s}}$ is the conditional mean and thus we have
\begin{align}
&{\mathbb{E}}\{({\bf{s}}-{\bf{\bar s}})({\bf{\bar s}}-{\bf{\hat s}})^{\rm{H}} \}={\bf{0}}
\end{align}based on which the right hand side of (\ref{covariance_matrix}) also equals to
\begin{align}
\label{covariance_matrix_a}
&\mathbb{E}\{({\bf{s}}-{\bf{\bar s}})({\bf{s}}-{\bf{\bar s}})^{\rm{H}}\}+
\mathbb{E}\{({\bf{\bar s}}-{\bf{\hat s}})({\bf{\bar s}}-{\bf{\hat s}})^{\rm{H}}\} \nonumber \\ &=\mathbb{E}\{({\bf{s}}-{\bf{\bar s}}+{\bf{\bar s}}-{\bf{\hat s}})({\bf{s}}-{\bf{\bar s}}+{\bf{\bar s}}-{\bf{\hat s}})^{\rm{H}}\}\nonumber \\
&=\mathbb{E}\{({\bf{s}}-{\bf{\hat s}})({\bf{s}}-{\bf{\hat s}})^{\rm{H}}\}.
\end{align}
Substituting (\ref{covariance_matrix_a}) into (\ref{covariance_matrix}), we have
\begin{align}
{\boldsymbol{\rm{Cov}}}({\bf{s}}|{\bf{y}},{\bf{y}}_1,{\bf{y}}_2)\preceq \mathbb{E}\{({\bf{s}}-{\bf{\hat s}})({\bf{s}}-{\bf{\hat s}})^{\rm{H}}\}={\boldsymbol { \Phi}}_{\rm{MSE}}({\bf{F}},{\bf{P}}).
\end{align}It is also well-known that with fixed covariance matrix, Gaussian distribution has the maximum entropy. Therefore, it is concluded that
\begin{align}
{\textsl{H}}({\bf{s}}|{\bf{y}},{\bf{y}}_1,{\bf{y}}_2)&\le \mathbb{E}\{{\rm{log}}|\pi e {\boldsymbol{\rm{Cov}}}({\bf{s}}|{\bf{y}},{\bf{y}}_1,{\bf{y}}_2)|\}\nonumber \\
& \le  {\rm{log}}|\pi e {\boldsymbol { \Phi}}_{\rm{MSE}}({\bf{F}},{\bf{P}})|,
\end{align}based on which, an lower bound of the capacity (\ref{capacity_equ}) is \begin{align}
&\textsl{I}({\bf{y}};{\bf{s}}|{\bf{y}}_1,{\bf{y}}_2)\nonumber \\
&\ge{\textsl{H}}({\bf{s}})-{\rm{log}}|\pi e{\boldsymbol {\Phi}}_{\rm{MSE}}({\bf{
F}},{\bf{P}}) |= -{\rm{log}}|{\boldsymbol {\Phi}}_{\rm{MSE}}({\bf{
F}},{\bf{P}}) |.
\end{align}The final equality comes from the fact ${\bf{R}}_{\bf{s}}={\bf{I}}$.

\section{Proof of Conclusion 1}
\label{Appedix:3}

As ${\boldsymbol g}(\bullet)$ is a matrix monotonically decreasing function, it can be proven that for the optimal solution the power constraint is always active, i.e., ${\rm{Tr}}({\bf{X}}{\bf{X}}^{\rm{H}})=P$ \cite{Jorswieck07}. As a result, we have the following relationship.
\begin{align}
\label{eta_x}
\eta_{x}&=\alpha{\rm{Tr}}({\bf{X}}{\bf{X}}^{\rm{H}}{\boldsymbol \Psi})+\sigma_{n}^2 \nonumber \\ &=\alpha {\rm{Tr}}({\bf{X}}{\bf{X}}^{\rm{H}}{\boldsymbol \Psi})+\sigma_{n}^2
\underbrace{{\rm{Tr}}({\bf{X}}{\bf{X}}^{\rm{H}})/P}_{=1} \nonumber \\ &={\rm{Tr}}({\bf{X}}{\bf{X}}^{\rm{H}}(\alpha P{\boldsymbol
\Psi}+\sigma_{n}^2{\bf{I}}))/P.
\end{align}From (\ref{eta_x}), the constraint of the optimization problem (\ref{Opt_X}) equals to
\begin{align}
&{\rm{Tr}}({\bf{X}}{\bf{X}}^{\rm{H}})={\rm{Tr}}[{\bf{X}}{\bf{X}}^{\rm{H}}(\alpha P{\boldsymbol
\Psi}+\sigma_{n}^2{\bf{I}})]/\eta_{x}=P,
\end{align}based on which the optimization problem (\ref{Opt_X}) is equivalent to
\begin{align}
\label{opt:app_3_1}
& \min_{{\bf{X}}} \ \ \  {\boldsymbol g}\left({\bf{X}}^{\rm{H}}{\bf{H}}^{\rm{H}}{\bf{H}}\frac{1}{\eta_x}{\bf{X}}\right) \nonumber \\
& \ {\rm{s.t.}} \ \ \ \ {\rm{Tr}}[{\bf{X}}{\bf{X}}^{\rm{H}}(\alpha P{\boldsymbol
\Psi}+\sigma_{n}^2{\bf{I}})]/\eta_{x}=P.
\end{align}Then, defining a new variable
\begin{align}
\label{App_equ_X}
{\bf{\tilde X}}={1}/{\sqrt{\eta_x}}(\alpha P{\boldsymbol
\Psi}+\sigma_{n}^2{\bf{I}})^{1/2}{\bf{X}},
\end{align}the optimization problem (\ref{opt:app_3_1}) is further reformulated as
\begin{align}
& \min_{{\bf{\tilde X}}} \ \ \  {\boldsymbol g}\left({\bf{\tilde X}}^{\rm{H}}(\alpha P{\boldsymbol \Psi}+{\sigma}_{n}^2{\bf{I}})^{-1/2}{\bf{H}}^{\rm{H}}{\bf{H}}(\alpha P{\boldsymbol \Psi}+{\sigma}_{n}^2{\bf{I}})^{-1/2}{\bf{\tilde X}}\right) \nonumber \\
& \ {\rm{s.t.}} \ \ \ \ {\rm{Tr}}({\bf{\tilde X}}{\bf{\tilde X}}^{\rm{H}})= P.
\end{align}

For any given ${\bf{\tilde X}}$, based on the following singular decompositions
\begin{align}
& {\bf{H}}(\alpha P{\boldsymbol \Psi}+{\sigma}_{n}^2{\bf{I}})^{-1/2}{\bf{\tilde X}}={\bf{V}}_{\boldsymbol \Xi}{\boldsymbol \Lambda}_{\boldsymbol \Xi}{\bf{U}}_{\boldsymbol \Xi}^{\rm{H}} \ \ {\text{with}} \ \ {\boldsymbol \Lambda}_{\boldsymbol \Xi} \searrow \nonumber \\
& {\bf{H}}(\alpha P{\boldsymbol \Psi}+{\sigma}_{n}^2{\bf{I}})^{-1/2}={\bf{U}}_{\bf{H}}{\boldsymbol \Lambda}_{\bf{H}}{\bf{V}}_{\bf{H}}^{\rm{H}} \ \ {\text{with}} \ \ {\boldsymbol \Lambda}_{\bf{H}} \searrow,
\end{align} there exists a matrix ${\bf{\bar X}}$ satisfying
\begin{align}
\label{X_OPT_STR}
{\bf{\bar X}}&={\bf{V}}_{\bf{H}}{\boldsymbol \Lambda}_{\bf{X}}{\bf{U}}_{\boldsymbol{\Xi}}^{\rm{H}} \\
\ {\rm{with}} \ \ & 1/b{\boldsymbol \Lambda}_{\bf{H}}{\boldsymbol \Lambda}_{\bf{X}}={\boldsymbol \Lambda}_{\boldsymbol \Xi} \ \searrow
\end{align}where ${\boldsymbol \Lambda}_{\bf{X}}$ is a diagonal matrix with the same rank as ${\boldsymbol \Lambda}_{\boldsymbol \Xi}$ and $b$ is a scalar which makes ${\rm{Tr}}({\bf{\bar X}}{\bf{\bar X}}^{\rm{H}})= P$ hold. Based on Lemma 12 in \cite{Palomar03}, the following inequality holds
\begin{align}
&{\bf{\bar X}}^{\rm{H}}(\alpha P{\boldsymbol \Psi}+{\sigma}_{n}^2{\bf{I}})^{-1/2}{\bf{H}}^{\rm{H}}{\bf{H}}(\alpha P{\boldsymbol \Psi}+{\sigma}_{n}^2{\bf{I}})^{-1/2}{\bf{\bar X}}\nonumber \\
\succeq &{\bf{\tilde X}}^{\rm{H}}(\alpha P{\boldsymbol \Psi}+{\sigma}_{n}^2{\bf{I}})^{-1/2}{\bf{H}}^{\rm{H}}{\bf{H}}(\alpha P{\boldsymbol \Psi}+{\sigma}_{n}^2{\bf{I}})^{-1/2}{\bf{\tilde X}}.
\end{align}Together with the fact that ${\boldsymbol g}(\bullet)$ is a matrix monotonically decreasing function, the following inequality holds
\begin{align}
&{\boldsymbol g}({\bf{\bar X}}^{\rm{H}}(\alpha P{\boldsymbol \Psi}+{\sigma}_{n}^2{\bf{I}})^{-1/2}{\bf{H}}^{\rm{H}}{\bf{H}}(\alpha P{\boldsymbol \Psi}+{\sigma}_{n}^2{\bf{I}})^{-1/2}{\bf{\bar X}})
\le \nonumber \\&{\boldsymbol g}({\bf{\tilde X}}^{\rm{H}}(\alpha P{\boldsymbol \Psi}+{\sigma}_{n}^2{\bf{I}})^{-1/2}{\bf{H}}^{\rm{H}}{\bf{H}}(\alpha P{\boldsymbol \Psi}+{\sigma}_{n}^2{\bf{I}})^{-1/2}{\bf{\tilde X}}).
\end{align}Therefore, it is concluded that the optimal ${\bf{\tilde X}}$ has the structure given by (\ref{X_OPT_STR}). Furthermore, based on the definition of  ${\bf{\tilde X}}$ (\ref{App_equ_X}), the optimal ${\bf{X}}$ has the following structure
\begin{align}
\label{X_structure}
{\bf{X}}_{\rm{opt}}=\sqrt{\eta_x}(\alpha P{\boldsymbol \Psi}+{\sigma}_{n}^2{\bf{I}})^{-1/2}{\bf{V}}_{\bf{H}}{\boldsymbol \Lambda}_{\bf{X}}{\bf{U}}_{\boldsymbol {\Xi}}^{\rm{H}}
\end{align}where ${\boldsymbol \Lambda}_{\bf{X}}$ is a diagonal matrix. In (\ref{X_structure}) $\eta_x$ is unknown either. In order to solve
$\eta_x$, substitute the structure of ${\bf{X}}$ in (\ref{X_structure}) into the definition of $\eta_x$ in
(\ref{Opt_X}), and then we get the following equation
\begin{align}
\eta_x&={\rm{Tr}}({\bf{X}}{\bf{X}}^{\rm{H}}{\boldsymbol
\Psi})\alpha+\sigma_{n}^2 \nonumber \\
&=\eta_x\alpha{\rm{Tr}}[{\bf{V}}_{{\bf{X}}}^{\rm{H}}
(\alpha P{\boldsymbol
\Psi}+\sigma_{n}^2{\bf{I}})^{-1/2}{\boldsymbol
\Psi}(\alpha P{\boldsymbol
\Psi}+\sigma_{n}^2{\bf{I}})^{-1/2}\nonumber \\
&\times{\bf{V}}_{{\bf{X}}}{\boldsymbol{\Lambda}}_{{\bf{ X}}}{\boldsymbol{\Lambda}}_{{\bf{ X}}}^{\rm{T}}]+\sigma_{n}^2.
\end{align}This is a simple linear function of ${\eta}_x$, and $\eta_x$ can be easily solved to be
\begin{align}
\eta_x&={\sigma_{n}^2}/\{1-\alpha{\rm{Tr}}[{\bf{V}}_{{\bf{X}}}^{\rm{H}}(\alpha P{\boldsymbol
\Psi}+\sigma_{n}^2{\bf{I}})^{-1/2}{\boldsymbol
\Psi}(\alpha P{\boldsymbol
\Psi}+\sigma_{n}^2{\bf{I}})^{-1/2}\nonumber \\
& \times {\bf{V}}_{{\bf{X}}}{\boldsymbol{\Lambda}}_{{\bf{ X}}}{\boldsymbol{\Lambda}}_{{\bf{ X}}}^{\rm{T}}]\}.
\end{align}

\section{Proof of Conclusion 2}
\label{Appedix:4}
The objective function in (\ref{objective}) can be reformulated as
\begin{align}
&{\rm{log}}|{\bf{A}}^{\rm{H}}({\bf{X}}^{\rm{H}}{\bf{H}}^{\rm{H}}
{\bf{H}}{\bf{X}}/\eta_x+{\bf{I}})^{-1}{\bf{A}}+{\bf{N}}|\nonumber \\
=&{\rm{log}}|{\bf{N}}||{\bf{A}}^{\rm{H}}({\bf{X}}^{\rm{H}}{\bf{H}}^{\rm{H}}
{\bf{H}}{\bf{X}}/\eta_x+{\bf{I}})^{-1}{\bf{A}}{\bf{N}}^{-1}
+{\bf{I}}| \nonumber \\
=&{\rm{log}}|{\bf{N}}|+{\rm{log}}|({\bf{X}}^{\rm{H}}{\bf{H}}^{\rm{H}}
{\bf{H}}{\bf{X}}/\eta_x+{\bf{I}})^{-1}{\bf{A}}{\bf{N}}^{-1}{\bf{A}}^{\rm{H}}
+{\bf{I}}| \nonumber \\
=&{\rm{log}}|{\bf{N}}|+{\rm{log}}|
{\bf{A}}{\bf{N}}^{-1}{\bf{A}}^{\rm{H}}+({\bf{X}}^{\rm{H}}{\bf{H}}^{\rm{H}}
{\bf{H}}{\bf{X}}/\eta_x+{\bf{I}})|\nonumber \\
&-{\rm{log}}|({\bf{X}}^{\rm{H}}{\bf{H}}^{\rm{H}}
{\bf{H}}{\bf{X}}/\eta_x+{\bf{I}})|
\end{align}where the second equality is based on the fact that $|{\boldsymbol {A}}{\boldsymbol {B}}+{\bf{I}}|=|{\boldsymbol {B}}{\boldsymbol {A}}+{\bf{I}}|$.
Using the matrix inequality that for two positive semi-definite matrices ${\boldsymbol M}$ and ${\boldsymbol N}$ i.e., $|{\boldsymbol M}+{\boldsymbol N}|\ge \prod (\lambda_i({\boldsymbol M})+\lambda_i({\boldsymbol N})) $ \cite{Marshall79}, we directly have
\begin{align}
\label{case_b}
& {\rm{log}}|{\bf{A}}^{\rm{H}}({\bf{X}}^{\rm{H}}{\bf{H}}^{\rm{H}}
{\bf{H}}{\bf{X}}/\eta_x+{\bf{I}})^{-1}{\bf{A}}+{\bf{N}}|
  \ge\nonumber \\&{\rm{log}}|{\bf{N}}|+ \sum_i{\rm{log}}[\lambda_i({\bf{A}}{\bf{N}}^{-1}{\bf{A}}^{\rm{H}})
+\lambda_i({\bf{X}}^{\rm{H}}{\bf{H}}^{\rm{H}}
{\bf{H}}{\bf{X}}/\eta_x+{\bf{I}})]\nonumber \\
&-\sum_i{\rm{log}}[\lambda_i({\bf{X}}^{\rm{H}}{\bf{H}}^{\rm{H}}
{\bf{H}}{\bf{X}}/\eta_x+{\bf{I}})].
\end{align}Together with the optimal structure given by \textbf{Conclusion 1}, in order to make the equality in (\ref{case_b}) hold
the following equation holds
\begin{align}
{\bf{U}}_{\boldsymbol{\Xi}}={\bf{U}}_{\bf{ANA}}.
\end{align} In light of the fact that power is never loaded to the eigenchannels with zero magnitudes \cite{Palomar03}, the diagonal matrix ${\boldsymbol \Lambda}_{\bf{x}}$ has the following structure
\begin{align}
{\boldsymbol \Lambda}_{\bf{x}}=
\left[ {\begin{array}{*{20}c}
   {{\boldsymbol {\tilde \Lambda}}_{\bf{x}}} & {{\bf{0}}}  \\
   {{\bf{0}}} & {{\bf{0}}}  \\
\end{array}} \right],
\end{align}where ${\boldsymbol {\tilde \Lambda}}_{\bf{x}}$ is a $N_X \times N_X$ diagonal matrix and $N_X=\min\{{\rm{Rank}}({\bf{H}}^{\rm{H}}{\bf{H}}), {\rm{Rank}}({\bf{A}}{\bf{A}}^{\rm{H}}) \}$.


\end{document}